\documentstyle[prl,aps]{revtex}
\begin{document}
\preprint{machin}
%


\title{Canonically invariant formulation of Langevin and Fokker-Planck equations}

\author{Olivier C\'epas and Jorge Kurchan\\
Laboratoire de Physique Th\'eorique 
{\sc enslapp}\thanks{URA 14-36 du CNRS,
associ\'ee \`a l'E.N.S. de Lyon, 
et \`a l'Universit\'e de Savoie}\\
ENSLyon,
46 All\'ee d'Italie,\\
F-69364 Lyon CEDEX 07, FRANCE\\
}
\date{June 1997}

\maketitle

\begin{abstract}

We present a canonically invariant form for the generalized Langevin and 
Fokker-Planck equations.  We discuss the role of constants of motion, and
the construction of conservative stochastic processes.

\end{abstract}

\vskip 1cm
\rightline{{\small E}N{\large S}{\Large L}{\large A}P{\small P}-L-654/97}
\vskip 1cm



Consider  the usual  Langevin equation:

\begin{equation}
 m\ddot{q}_i = -\frac{\partial V}{\partial q_i}- \dot{q_i} + \xi_i(t) 
\label{Langevin1}
\end{equation} 
with $\xi_i(t)$  Gaussian white noises  
$<\xi_i(t)\xi_j(t')>= 2  T \delta_{ij}\delta(t-t')$, and
 $T$ the  temperature of the thermic bath.
Rewriting this as a  set of phase-space equations:
\begin{eqnarray}
 \left \{ \begin{array}{cccc} 
          & \dot{q}_i  &=& \frac{p_i}{m}   \\
  & \dot{p}_i  &=& -\frac{\partial V}{\partial q_i}-   \frac{p_i}{m} + \xi_i(t)
                        \end{array} \right.
\label{Langevin2}
\end{eqnarray}
we notice two things. First, the form of the first equation is restricted to
a hamiltonian $H$ of the form $H=\sum_ip_i^2/m + V(q)$. Second, the interaction with
the bath has introduced  an 
assymmetry in the treatment of coordinates and momenta.
We shall in what follows formulate the  Langevin and Fokker-Planck processes in
a way that treats all phase-space variables on an equal footing.

In general, Langevin  equations 
can be motivated \cite{Zwanzig}\cite{Hanggi} by considering the system with 
Hamiltonian $H$,
 coupled to an infinite set of harmonic oscillators with  random phases at some initial
time and energies given by equipartition at temperature $T$. Upon solving for the oscillators,
and reinjecting their dependence on the equation of motion, one gets a Langevin equation which
can be made Markovian by a suitable choice of distribution of the  oscillators' frequencies.

Actually, (\ref{Langevin1},\ref{Langevin2}) are associated with a 
particular coupling  of the form:
\begin{equation}
H_{coup} =\sum_i q_i \left [ \sum_{a=1}^{N} A^i_a y_a^i \right]
\end{equation}
where $y_a^i$ are the coordinates of the oscillators of frequencies $\omega_a^i$.

In order to obtain a canonically invariant generalization, one can repeat the exercise with
 a coupling
 with the `bath' of 
oscillators of a more general form:
\begin{equation}
H_{coup} =\sum_i \sum_{a=1}^N [A^i_a G^1_{i}({\bbox q},{\bbox p}) y_a^i+
B^i_a G^2_{i}({\bbox q},{\bbox p}) {\dot y}_a^i]
\label{cosa}
\end{equation}
Performing the usual steps, one arrives at the following Langevin equation, valid 
for  any phase-space variable $A({\bbox q},{\bbox p})$, in particular the coordinates and 
momenta $q_i,p_i$:
\begin{equation} 
 \dot{A} = \kappa \{A,H\}+ \sum_j \{A,G_j\}(\xi_j(t)+\{G_j,H\})
\label{Langevin3}
\end{equation}
We have explicitated an inverse  time-constant $\kappa$. 
Here  
$\{A,B\} = \sum_i \big(\frac{\partial A}
{\partial q_i} \frac{\partial B}{\partial p_i} - \frac{\partial A}{\partial p_i}
\frac{\partial B}{\partial q_i} \big)$ are the Poisson brackets.

The $G_i({\bbox p},{\bbox q})$ ($i=1,...,R$) are $R$ arbitrary phase-space functions, originally
controlling the manner of the coupling to the bath as in Eq. (\ref{cosa}). The noise
is white and gaussian as before.

In fact, we do not need to go  into the details of the derivation, because it
will be shown in what follows that this equation is a {\em bona-fide} Langevin equation
in that it leads to the canonical distribution at temperature $T$.
The definition of Equation  (\ref{Langevin3}) is completed by specifying that
it should be understood {\em in the Stratonovitch sense}: in a discretized form
all phase-space functions in the right hand side have to be evaluated as an average of their
values in the previous and the incremented time.  

Indeed, one can adopt \textit{It\^o's convention} so the r.h.s. is evaluated in the previous time, and
the equation now reads:
\begin{equation} 
\dot{A} =  \kappa \{A,H\}+\sum_j \{A,G_j\}(\xi_j(t)+\{G_j,H\})+  T\{G_j,\{G_j,A\}\}
\label{Langevino}
\end{equation}

Two particular cases are $G_i=-q_i \; \forall \; i \; ; \; \kappa=1$ ,  which yields  
(\ref{Langevin1},\ref{Langevin2}), and $G_i=p_i \; \forall \; i \; ; \; \kappa=0$ which yields the
 {\em massless} version
of (\ref{Langevin1}). Note that in general the Poisson brackets between
the $G_i$ need not vanish, in which case the equations (\ref{Langevin3})
cannot be taken
through a canonical transformation to the form (\ref{Langevin1}). 

Let us now turn to the (Fokker-Planck) equation satisfied by the probability
distribution $P({\bbox q},{\bbox p},t)$. It is a simple exercise (see \cite{Risken} or \cite{Zinn-Justin})
to obtain this directly from equation (\ref{Langevin3}) or (\ref{Langevino}).
The result  is:

\begin{equation} 
\frac{\partial P}{\partial t}+ \kappa \{P,H\} = \sum_j\{G_j,\{G_j,H\}P+T\{G_j,P\}\} 
\label{Fokker}
\end{equation}
It is now clear that $P=\exp(-H/T)$ is a stationary solution of 
(\ref{Fokker}).
Again, with the choice $G_i=-q_i \; \forall \; i \; ; \; \kappa=1$  we obtain the Kramer's equation.
 If instead we make $G_i=p_i\; \forall \; i \; ; \; \kappa=0 $ we obtain the
 usual Fokker-Planck equation for diffusion without
inertia.

By writing $\langle A \rangle(t) = 
\int d{\bbox q}\; d{\bbox p} A({\bbox q},{\bbox p}) P({\bbox q},{\bbox p},t)$ we obtain for
 the evolution of
the average of an observable $A({\bbox q},{\bbox p})$:

\begin{equation} \frac{d \langle A \rangle(t)}{dt} = 
\langle \{A,H\}\rangle -\sum_i \langle \{G_i,A\}\{G_i,H\} \rangle - 
T \sum_i   \langle \{G_i,\{A,G_i\}\} \rangle 
\label{average}
\end{equation}

All three equations (\ref{Langevin3}), (\ref{Fokker}) and (\ref{average})
are canonically invariant in form: a canonical transformation of variables
is obtained directly by transforming $H$ and the $G_i$. Furthermore,
 the pure
Hamiltonian  term and the bath-coupling terms are   explicitly separated.

{\bf Equilibration}

In order to study the equilibration properties, let us define an $\mathcal H$-function as \cite{Kubo}\cite{De Groot} :
\begin{equation}
\mathcal H (t) = \int d{\bbox q}\; d{\bbox p} P({\bbox q},{\bbox p},t)  \big(T 
\ln P({\bbox q},{\bbox p},t) + H({\bbox q},{\bbox p}) \big)
\end{equation}
which cannot increase, since:
\begin{equation}
\dot{\cal H}(t) = -\sum_i  \int  d{\bbox q}\; d{\bbox p}
\frac{(\{G_i,H\}P+T\{G_i,P\})^2}{P} \leq0
\end{equation}
Note that only the bath-terms contribute.

If the equilibrium measure exists, ${\cal H}$ is bounded from below, and we have that
\begin{equation} 
\{G_i,H\}P+T\{G_i,P\} \rightarrow 0 \;\;\; \forall \; i
\label{equi}
\end{equation}
If we parametrize $P({\bbox q},{\bbox p},t)$ as:
\begin{equation}
P({\bbox q},{\bbox p},t)=Q({\bbox q},{\bbox p},t) \exp(-H/T)
\end{equation}
the limit (\ref{equi}) implies that, once stationarity is achieved :
\begin{equation}
\{G_i,Q\}=0 \;\;\; \forall \; i
\label{machin}
\end{equation}
Using this equation, we have that $\dot{Q}=\{H,Q\}$. Since at stationarity (\ref{machin})
has to be valid at all times, we obtain the necessary conditions:

\begin{equation} 
\{G_i,Q\}=0 \;\;\; ; \;\;\;
\{G_i,\{H,Q\}\}=0\;\;\; ; \;\;\;
\{G_i,\{H,\{H,Q\}\}\}=0\;\;\; ; \;\;\; \dots
\end{equation}

In the usual Langevin case (\ref{Langevin1},\ref{Langevin2}), $G_i=x_i$, $H=\sum_i p_i^2/2m +V(x)$
and the first two sets of equations suffice to prove that $Q={\mbox {constant}}$ is the 
only stationary solution.

{\bf Constants of Motion}

 Suppose the Hamiltonian has some constants of motion  $\{H,K_a\}=0$.
Depending on the choice of $G_i$, these constants will be preserved or not by the 
coupling with the bath. Indeed, Eq. (\ref{Langevin3}) implies:
\begin{equation}
\frac{dK_a}{dt} =\sum_j  \{K_a,G_j\}(\xi_j(t)+ \{G_j,H\})
\end{equation}
The evolution of $K_a$ is then purely dictated by the heat-bath, and will be `slow' in
the small noise limit.

If we wish to  construct a $K_a$-preserving noisy dynamics we have to choose
the $G_i$ such that $\{K_a,G_i\}=0 \; \forall \; i$.

If, on the other hand, we couple the system to the bath through some constants of
motion, that is $G_i=K_i$, the Langevin dynamics for any $A({\bbox q},{\bbox p})$ becomes: 
\begin{equation}
 \dot{A} = \kappa \{A,H\} + \sum_j  \{A,K_j\}\xi_j(t)
\end{equation} 
which expresses the fact that the system receives random kicks in the direction generated
by the $K_j$.

An extreme and rather amusing form of this is the case in which we put a single $G=H$ and
$\kappa=0$.
We then have 
\begin{equation}
 \dot{A} =   \{A,H\}\xi(t)
\end{equation} 
and the associated Fokker-Planck equation:
\begin{equation}
\frac{\partial P}{\partial t} = T \{H,\{H,P\}\} 
\end{equation}
The system  diffuses back an forth along its classical trajectories. The probability distribution
tends for long times to the smallest invariant structure compatible with the original distribution,
 and
the  entropy $\int d{\bbox q} d{\bbox p} P({\bbox q},{\bbox p},t) \ln P({\bbox q},{\bbox p},t)$
becomes stationary.
 
{\bf Motion within a Group}

Another simple application is the construction of a heat-bath dynamics on a group.
Suppose the Hamiltonian is constructed in terms of the generators $L_i$ of a group,  satisfying
$\{L_i,L_j\} = C_{ijl}L_l$.
Then, 
\begin{equation}
 \dot{L}_i = \{L_i,H\}+C_{ijl}L_l(\xi_j(t)+ \{L_j,H\}) 
=C_{ijl} (\omega_j + \xi_j) L_l + C_{ijl} C_{jsr} \omega_s  L_r  L_l
\end{equation} 
where the `angular velocities' are defined as $\omega_i=\partial H({\bbox L})/\partial L_i$.
The group invariant $\sum_i L_i^2$ is clearly a constant of motion.

\vspace{2cm}

In summary, we have presented a manifestly canonical-invariant form of the langevin
and Fokker-Planck equations. Within this formulation several questions originating from 
the underlying classical mechanics  become more transparent.

\end{document}